\documentclass[10pt,twocolumn]{article}

\usepackage{graphicx}
\usepackage{amssymb}
\usepackage{amsmath}
\usepackage{bm}
\usepackage{color}
\usepackage{authblk}
\usepackage{ar}
\usepackage{lineno}

\title{Added costs of insect-scale flapping flight in unsteady airflows}
\author[1,*]{Dmitry Kolomenskiy}
\author[2,3,*]{Sridhar Ravi}
\author[1]{Taku Takabayashi}
\author[1]{Teruaki Ikeda}
\author[1]{\\Kohei Ueyama}
\author[4]{Thomas Engels}
\author[2]{Alex Fisher}
\author[5]{Hiroto Tanaka}
\author[6]{\\Kai Schneider}
\author[4]{J\"orn Sesterhenn}
\author[1,7]{Hao Liu}
\affil[1]{Graduate School of Engineering, Chiba University, Chiba, Japan}
\affil[2]{School of Aerospace Mechanical and Manufacturing Engineering, RMIT University, Melbourne, Australia}
\affil[3]{Department of Neurobiology, University of Bielefeld, Bielefeld, Germany}
\affil[4]{ISTA, Technische Universit\"at Berlin, Berlin, Germany}
\affil[5]{Department of Mechanical Engineering, School of Engineering, Tokyo Institute of Technology, Tokyo, Japan}
\affil[6]{I2M - UMR 7373 - CNRS, Centre de Math\'ematiques et d'Informatique,
Aix-Marseille Universit\'e, Marseille, France}
\affil[7]{Shanghai-Jiao Tong University and Chiba University International Cooperative Research Centre (SJTU-CU ICRC),
Shanghai, China}
\affil[*]{Equal contributions}

\begin{document}


\twocolumn[
\begin{@twocolumnfalse}
\maketitle
\begin{abstract}
The aerial environment in the operating domain of small-scale natural and
artificial flapping wing fliers is highly complex, unsteady and generally turbulent. 
Considering flapping flight in an unsteady wind environment with
a periodically varying lateral velocity component, we show that body
rotations experienced by flapping wing fliers result
in the reorientation of the aerodynamic force vector that can render a substantial
cumulative deficit in the vertical force.
We derive quantitative estimates of the body roll amplitude
and the related energetic requirements to maintain the weight support in free flight under such conditions.
We conduct force measurements of a miniature hummingbird-inspired robotic flapper
and numerical simulations of a bumblebee.
In both cases, we demonstrate the loss of weight support due to body roll rotations.
Using semi-restrained flight
measurements, we demonstrate
the increased power requirements to maintain altitude in unsteady winds, achieved by increasing the flapping frequency.
Flapping fliers may increase their flapping frequency as well as the stroke amplitude to
produce the required increase in
aerodynamic force, both of these two types of compensatory control requiring additional energetic
cost.
We analyze the existing data from experiments on animals flying in von K\'arm\'an streets
and find reasonable agreement with the proposed theoretical model.
\end{abstract}
\end{@twocolumnfalse}
]

~

\section{Introduction}

Flying animals and micro aerial vehicles (MAVs) typically operate in highly unsteady turbulent winds that may influence the energetic cost of flight.
Time variation of the prevailing wind speed and direction, time-periodic fluctuation in wakes behind various objects,
and fully developed turbulence at smaller scales are some typical examples of wind unsteadiness.
Our understanding of the influence of wind unsteadiness on flapping flight remains limited and
the mechanisms that potentially lead to increased energetic costs remain largely unexplored. 

Volant insects, small birds and miniature flying vehicles are compact and lightweight, thus possessing low inertia.
From the flight dynamics standpoint, this renders them particularly sensitive to wind unsteadiness \cite{Shyy_etal_2015_proca,Liu_etal_2016_rspb}.
Free flight measurements made on bumblebees and hawkmoths reveal that in unsteady winds animals
frequently change body orientation as a consequence of active and passive interactions with the airflow \cite{Ravi_etal_2013_jeb,Ortega-Jimenez_etal_2013_jeb}.
In the wake of a circular cylinder, bumblebees experience variations in body roll of up to 15~deg while hawkmoths in
nominally similar conditions undergo nearly 23~deg \cite{Ortega-Jimenez_etal_2013_jeb}.
Other insects such as stalk-eye flies, honeybees also experience significant body rotations, when subjected to strong gusts \cite{Vance_etal_2013_bb}.
Similar observations have been made on hummingbirds flying in unsteady winds and
increased metabolic rates during flight in such flow conditions have been reported \cite{Ortega-Jimenez_etal_2014_procb}.

Few studies have also analysed the influence of wind unsteadiness on the mean aerodynamic force production of flapping wings. At the scale of bumblebees,
high fidelity numerical simulations of a tethered bumblebee revealed that even high levels of freestream turbulence (intensity $> 30\%$)
that are commonly encountered in atmospheric winds do not deteriorate aerodynamic performance \cite{Engels_etal_2016_prl}.
However, the freestream turbulence induced large instantaneous variation in forces and created body torques that would have challenged flight control \cite{Engels_etal_2016_prl}.
Tests performed on a robotic flapping wing operating at larger Reynolds numbers ($Re = 50,000$) in different levels of fully developed freestream turbulence also revealed that
the fluctuations in the aerodynamic forces increased with increasing levels of turbulence \cite{Fisher_etal_2016_bb}.
The magnitude of the fluctuations in turbulent winds in relation
to the forces produced in quiescent conditions at commensurate mean airspeeds decreased with increasing reduced flapping frequency \cite{Fisher_etal_2016_bb}.
This provides some indication that flapping flight may offer some advantageous aerodynamic gust mitigating properties over conventional static wings with
streamlined airfoils, which can be extremely sensitive to the free stream disturbance \cite{Mueller_etal_1983_expfluids}. However further systematic investigations are necessary to obtain a better understanding the aerodynamic interactions between flow unsteadiness and flapping wings.

Here we explore one of the factors that can increase energetic cost of flight in time-periodic unsteady flows: reorientation of the aerodynamic force vector.
We start with the fact that small-scale flapping wing fliers are prone to variation in body orientation \cite{Ravi_etal_2013_jeb,Ortega-Jimenez_etal_2013_jeb,Vance_etal_2013_bb,Ortega-Jimenez_etal_2014_procb}.
For flying insects and robots, the mean aerodynamic force vector is generally fixed with respect to the body \cite{Shyy_etal_2015_proca} and the
helicopter model of flight control is acceptable for manoeuvres at timescales sufficiently larger than the flapping period. As per this model, accelerations during manoeuvreing are
produced through stroke plane re-orientation \cite{Leishman_2006}. They render a non-zero component of the mean aerodynamic force vector in the desired direction.
A corollary is that when fliers change body orientation, actively or passively, the orientation of the total aerodynamic force vector is also likely to change.
The helicopter model has been successfully used to explain flight manoeuvres in fruitflies \cite{Muijres_etal_2015_jeb}, hawkmoths \cite{Greeter_Hedrick_2016_jeb}, bumblebees \cite{Ravi_etal_2013_jeb,Ravi_etal_2016_srep}, etc., though exceptions can occur when the aerodynamic force vector may be strongly decoupled from the body orientation.
Body rotations experienced inflight due to interactions with unsteady wind can induce continuous reorientation of the resultant aerodynamic force
vector with respect to the gravitational axis, which can in turn result in a cumulative reduction in altitude. Thus, in order to maintain altitude
in spite of body rotations, either due to self-initiated manoeuvres or due to wind-induced disturbance, overall increase in vertical force may be necessary.
Hence, reorientations induced by wind unsteadiness can be one of factors that may increase power requirements, as compared to straight and
level flight in quiescent conditions with no body rotations.

We focus on the effect of almost periodic roll oscillations of a flier induced by the vortex shedding from a vertical cylinder in otherwise uniform steady flow,
which is a canonical ecologically relevant aerodynamic perturbation \cite{Ravi_etal_2013_jeb,Ortega-Jimenez_etal_2013_jeb,Ortega-Jimenez_etal_2014_procb}.
Thus, in the remaining sections, ``unsteady wind" mainly refers to the velocity field in the wake behind a cylinder,
but similar considerations may apply to other degrees of freedom (DOF) and different kinds of perturbation.
The objective of this work is to evaluate the effect of such aerial environment
on the flier in terms of the roll angle dynamics, the aerodynamic vertical force production and the related power requirements.

In section~\ref{sec:theoretical_estimates}, we theoretically derive quantitative estimates of the roll angle variation and the added costs
imposed by unsteady wind on flapping flight.
Then, taking a bio-inspired approach and using a miniature mechanical flapping wing contrivance we compare the power required to maintain altitude during flight in quiescent
and unsteady wind conditions by performing a combination of force measurements in fully tethered (0-DOF), free to roll (1-DOF), and free to translate vertically and roll (2-DOF) 
conditions in section~\ref{sec:robotic_flapper_experiments}.
In section~\ref{sec:numerical_simulation}, we present high fidelity numerical simulations of a bumblebee at a lower Reynolds number.
We find a similar dynamic response, despite many morphological differences between the bumblebee and the robotic flapper.
Finally, the results obtained from experiments and numerical simulations are complied with measurements from other studies and put in perspective with the theoretical estimates
in section~\ref{sec:discussion}, and the main conclusions are summarized in section~\ref{sec:conclusions}.

\section{Theoretical estimates}\label{sec:theoretical_estimates}

\subsection{Dynamics of roll oscillations}\label{sec:roll_dynamics}

\begin{figure}
\begin{center}
\includegraphics[scale=1.1]{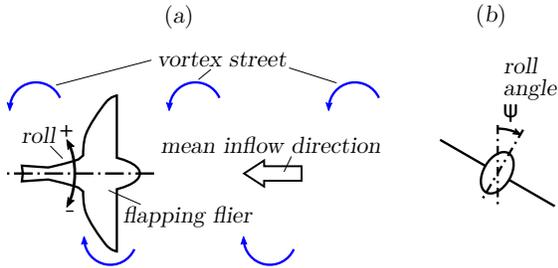}
\caption{\label{fig:overview} (\textit{a}) Schematic drawing of a model flapping flier in a vortex street.
The mean effective inflow velocity is the superposition of average wind and forward flight translation of the flier.
As the flier progresses through the vortex street, it experiences alternating lateral gusts
and rolls periodically about the longitudinal axis.
(\textit{b}) Front projection view along the roll axis, visualizing the roll angle $\psi$.}
\end{center}
\end{figure}

Body posture of a flapping flier in free flight can vary in time.
In this section, we derive closed-form expressions for the dynamics of body rotation
experienced by a miniature flapping wing flier in unsteady winds.
We focus on the rotation about the longitudinal axis, i.e., roll, which is particularly prone to instability in some animals \cite{Ravi_etal_2013_jeb}.
Similar considerations may apply to the dynamics of pitch rotations about the lateral axis.
Here, we consider a simplified model flapper with one degree of freedom, which is the roll angle $\psi$, subject to an external aerodynamic perturbation
while flying upstream in unsteady winds (figure~\ref{fig:overview}).
The aerodynamic perturbation of interest is a time-periodic lateral velocity
created, for instance, by a vertically oriented von K\'arm\'an street or by a series of lateral gusts, impinging on the flapper and producing unsteady
aerodynamic torques $\tau'$ about the roll axis of the flapper.

Under the assumptions of quasi-steady flapping wing aerodynamics,
it is possible to describe the body dynamics by only considering wingbeat averaged torques.
We use the quasi-steady approximations introduced in \cite{Cheng_Deng_2011_ieee} that were
based upon and subsequently validated in multiple studies on maneuvering and perturbation response in animals \cite{Cheng_etal_2011_jeb,Cheng_etal_2010_jeb,Hedrick_etal_2009_science}.
Hence, the general equation for the roll angle $\psi$ of the flapper has the form of a linear driven oscillator
\begin{equation}
I_{roll} \ddot{\psi} + \kappa_{fct} \rho R^5 f \dot{\psi} + l_p m_p g \psi = \tau',
\label{eq:eq_roll_motion}
\end{equation}
where dots stand for time derivatives.
The first term on the left hand side is the inertial torque, with $I_{roll}$ being the roll moment of inertia.
The second term is damping due to flapping counter torque, where the damping coefficient
depends on the wing length $R$, flapping frequency $f$, air density $\rho$, and a dimensionless coefficient that
can be estimated using quasi-steady approximations \cite{Hedrick_etal_2009_science} as
\begin{equation}
\kappa_{fct} = \overline{d \hat{\phi} / d \hat{t} ~c_F \sin{\alpha} } ~ \Phi \hat{r}_3^3 / \AR,
\label{eq:eq_kappa_fct}
\end{equation}
which is a function of the wingbeat amplitude $\Phi$, wing aspect ratio $\AR$,
nondimensional third moment of area $\hat{r}_3$,
the time average of the nondimensional wing angular velocity $d \hat{\phi} / d \hat{t}$,
aerodynamic resultant force coefficient $c_F$ and sine of the feathering angle $\alpha$.
Note that $d \hat{\phi} / d \hat{t}$, $c_F$ and $\alpha$ depend on time.
The third term in (\ref{eq:eq_roll_motion}) is a linearized pendulum restoring torque that
depends on the pendulum mass $m_p$, length $l_p$ and gravitational acceleration $g$.
The length $l_p$ is the distance between the center of mass and center of roll rotation, that vanishes in free flight,
but it can be non-zero in semi-tethered flights (e.g., all robotic flapper experiments conducted in this study).

The right hand side of (\ref{eq:eq_roll_motion}) is the aerodynamic torque $\tau'$ induced by the oncoming time-periodic unsteady flow (e.g., von K\'arm\'an street).
We only take the strongest Fourier mode into consideration which
corresponds to the characteristic frequency of the von K\'arm\'an street, i.e., we consider
\begin{equation}
\tau' = \tau'_a \cos{2 \pi f_{vk} t}, ~ \textrm{where}~ \tau'_a = \tau^*_a \rho R^5 f^2,
\label{eq:eq_torque}
\end{equation}
and $\tau^*_a$ is the dimensionless aerodynamic torque amplitude.

We define two dimensionless numbers to describe a periodic lateral velocity fluctuation:
\begin{itemize}
\item \emph{Normalized frequency} that relates the wing flapping frequency $f$ to the velocity fluctuation frequency $f_{vk}$,
\begin{equation}
\theta_{vk} = \frac{f}{f_{vk}};
\label{eq:thetavk}
\end{equation}
\item \emph{Turbulence intensity} that relates the root mean square (r.m.s.) of the lateral velocity fluctuations $W'$ to the wing length $R$ and the flapping frequency $f$,
\begin{equation}
Tu_w = \frac{W'}{R f}.
\label{eq:eq_turb_int}
\end{equation}
\end{itemize}

In order to relate the aerodynamic torque (\ref{eq:eq_torque}) to these two parameters of the velocity fluctuations, we expand
$\tau^*_a$ in power series of $Tu_w$,
\begin{equation}
\tau^*_a = \frac{\partial \tau^*_a}{\partial Tu_w} Tu_w + \frac{1}{2} \frac{\partial^2 \tau^*_a}{\partial Tu_w^2} Tu_w^2.
\label{eq:taua_series}
\end{equation}
The zeroth order coefficient in front of $(Tu_w)^0$ vanishes because $\tau^*_a$ tends to zero as $W'$ tends to zero.
Terms above second order do not appear since the aerodynamic stresses exerted on solid boundaries immersed in a fluid are at most quadratic in the velocity,
so (\ref{eq:taua_series}) is asymptotically exact.

From (\ref{eq:taua_series}) it follows that there are two distinct asymptotic regimes depending on $Tu_w$.
\begin{itemize}
\item If $Tu_w << 1$, the vortex street acts as a small perturbation that mainly changes the local angle of attack of the wings.
Therefore, the linear term is dominant.
This situation is common for most insects flying in mild turbulence since their flapping frequency is generally higher than the aerial disturbance,
except moths and butterflies that have significantly lower flapping frequency.
The roll torque is generated by the asymmetry of lift between left and right wing.
By analogy with the flapping counter torque (see the second term in (\ref{eq:eq_roll_motion})), the aerodynamic torque induced by the aerodynamic perturbation (e.g., von K\'arm\'an street) is approximated as
\begin{equation}
\tau'_a = c_{\tau1} \rho R^5 f \Omega,
\label{eq:eq_torque_lift}
\end{equation}
where $c_{\tau1}$ is a dimensionless coefficient similar to $\kappa_{fct}$, and the equivalent angular velocity is equal to $\Omega = W'/R$.
This yields an estimate for the first coefficient in the right-hand side of (\ref{eq:taua_series}),
\begin{equation}
\frac{\partial \tau^*_a}{\partial Tu_w} = c_{\tau1}.
\label{eq:torque_derivative_1}
\end{equation}

\item If $Tu_w >> 1$, the leading order contribution to the aerodynamic torque becomes due to quadratic drag (last term in (\ref{eq:taua_series})), if the center of drag force does not coincide with the center of mass as is the case for many biological and artificial fliers.
Hence, it can be approximated as
\begin{equation}
\tau'_a = c_{\tau 2} \frac{\rho W'^2 R^3}{2}.
\label{eq:eq_torque_drag}
\end{equation}
After non-dimensionalizing the torque with $\rho R^5 f^2$ and using (\ref{eq:eq_turb_int}), we obtain
\begin{equation}
\frac{\partial^2 \tau^*_a}{\partial Tu_w^2} = c_{\tau2}.
\label{eq:torque_derivative_2}
\end{equation}
\end{itemize}

The time-periodic solution of (\ref{eq:eq_roll_motion}) is
\begin{equation}
\psi(t) = \Psi \cos(2 \pi f_{vk} t - \xi),
\label{eq:theor_roll_angle}
\end{equation}
where
\begin{equation}
\Psi = \frac{\tau'_a}{\sqrt{(l_p m_p g - I_{roll}(2 \pi f_{vk})^2)^2+(2 \pi \kappa_{fct} \rho R^5 f f_{vk})^2}}
\label{eq:theor_roll_amp}
\end{equation}
and
\begin{equation}
\xi = \cot^{-1} \frac{l_p m_p g - I_{roll}(2 \pi f_{vk})^2}{2 \pi \kappa_{fct} \rho R^5 f f_{vk}}.
\label{eq:theor_roll_phshift}
\end{equation}
We are mainly interested in the roll amplitude $\Psi$, as it shall be shown later that $\Psi$ determines the mean vertical force deficit due to body rotations.
In free flight, the pendulum stability term in (\ref{eq:theor_roll_amp}) vanishes because
the center of rotation coincides with the center of mass, i.e., $l_p=0$.
Moreover, when the flapping frequency $f$ is large compared to the aerodynamic perturbation frequency $f_{vk}$,
the flapping counter torque term that contains $f$ dominates over the body inertia term in the denominator of (\ref{eq:theor_roll_amp}).
With these simplifications, after substituting $\tau_a'=\tau_a^* \rho R^5 f^2$, $f = \theta_{vk} f_{vk}$ and (\ref{eq:taua_series}) into (\ref{eq:theor_roll_amp}),
we obtain
\begin{equation}
\Psi \approx \frac{\theta_{vk}}{2 \pi} \left( \frac{c_{\tau1}}{\kappa_{fct}} Tu_w + \frac{1}{2} \frac{c_{\tau2}}{\kappa_{fct}} Tu_w^2 \right).
\label{eq:theor_roll_amp_approx}
\end{equation}
Equation (\ref{eq:theor_roll_amp_approx}) provides some useful insights into the body roll dynamics in unsteady wings.
For example, since $Tu_w \propto f^{-1}$ and $\theta_{vk} \propto f$, using (\ref{eq:taua_series}),
we obtain $\Psi \sim f^0$ when $f$ is large, and $\Psi \sim f^{-1}$ when $f$ is small.
Also, since $\theta_{vk} \propto f_{vk}^{-1}$, we see that the roll amplitude $\Psi$ becomes smaller as the perturbation frequency increases.
For the purpose of comparison with experiments, let us introduce the r.m.s. roll angle, which is equal to
\begin{equation}
\psi_{rms} = \Psi / \sqrt{2}
\label{eq:theor_roll_rms_approx}
\end{equation}
for the cosine wave (\ref{eq:theor_roll_angle}).

\subsection{Roll-induced mean vertical force deficit and mechanical power requirements}

In level flight, the mean aerodynamic force is equal in magnitude to the weight of the flier,
\begin{equation}
\overline{F}_{z0} = m g.
\label{eq:f0}
\end{equation}
Then if the flier rolls due to the aerodynamic disturbance while maintaining the wing kinematics, the vertical aerodynamic force in the global coordinate system (with $z$ axis opposite to the direction of gravity) depends on the roll angle $\psi(t)$:
\begin{equation}
F_z = \overline{F}_{z0} \cos \psi.
\label{eq:theor_fz_cos_roll}
\end{equation}
Here, we consider the situation when the flapper rolls about the mean inflow direction.
Similar analysis for the case of roll about the longitudinal body axis is discussed in supplementary material~S5.
For harmonic oscillations of $\psi(t)$ defined by (\ref{eq:theor_roll_angle}), time averaging of (\ref{eq:theor_fz_cos_roll}) over the vortex shedding period yields
\begin{equation}
\overline{F}_z = \overline{F}_{z0} J_0(\Psi),
\label{eq:theor_fz_avg}
\end{equation}
where $J_0$ is a Bessel function of the first kind.
Thus, the mean vertical force deficit, measured as the difference in vertical force produced in quiescent conditions and due to the body rotations in an unsteady flow, is equal to
\begin{equation}
\Delta \overline{F}_z = \overline{F}_{z0} - \overline{F}_{z} = \overline{F}_{z0} (1 - J_0(\Psi)) \approx \overline{F}_{z0} \frac{\Psi^2}{4}.
\label{eq:theor_fz_deficit}
\end{equation}
By substituting (\ref{eq:theor_roll_amp_approx}) for $\Psi$ in (\ref{eq:theor_fz_deficit}),
we obtain the following general relationship for the relative vertical force deficit for an arbitrary flapper in an unsteady flow:
\begin{equation}
\frac{\Delta \overline{F}_z}{\overline{F}_{z0}} = \left( \frac{\theta_{vk}}{4 \pi} \left( \frac{c_{\tau1}}{\kappa_{fct}} Tu_w + \frac{1}{2} \frac{c_{\tau2}}{\kappa_{fct}} Tu_w^2 \right) \right)^2.
\label{eq:theor_fz_deficit_2coef}
\end{equation}
This means that, in mild turbulence, $Tu_w << 1$, the relative force deficit does not depend on the flapping frequency $f$.
In strong turbulence, $Tu_w >> 1$, it scales like $R^{-4} f^{-2}$, which suggests that the absolute force $\Delta \overline{F}_z$ deficit should be approximately constant if the vertical force magnitude $\overline{F}_{z0}$ scales like $R^4 f^2$.

Finally, the mean vertical force deficit
due to body roll in unsteady flows translates into increased power requirements for level flight.
The required increase in mechanical power $\Delta \overline{P} \propto \Delta \overline{F}_z^\gamma$, $1 \le \gamma \le 1.5$, can be approximated to first order accuracy by
$\Delta \overline{P} = (\partial \overline{P} / \partial \overline{F}_{z}) \Delta \overline{F}_z$.
Using (\ref{eq:theor_fz_deficit_2coef}) and (\ref{eq:f0}),
we obtain the body mass specific mechanical power overhead cost due to body rotations,
\begin{equation}
\frac{\Delta \overline{P}}{m} = g \frac{\partial \overline{P}}{\partial \overline{F}_{z}} \left( \frac{\theta_{vk}}{4 \pi} \left( \frac{c_{\tau1}}{\kappa_{fct}} Tu_w + \frac{1}{2} \frac{c_{\tau2}}{\kappa_{fct}} Tu_w^2 \right) \right)^2.
\label{eq:theor_power}
\end{equation}

\section{Robotic flapper experiments}\label{sec:robotic_flapper_experiments}

\subsection{Fully tethered (0-DOF) and free to roll (1-DOF) measurements}\label{sec:tethered_experiment}

Wind tunnel experiments were conducted using a robotic flapper that converted rotational motion of a DC motor to reciprocating
motion of a pair of wings using a combination of gears and sliders.
A photographic image of the robotic flapper is shown in figure~\ref{fig:flapper_all}(\textit{a}), and its brief description can be found in supplementary material~S1.

\begin{figure*}
\begin{center}
\includegraphics[scale=0.9]{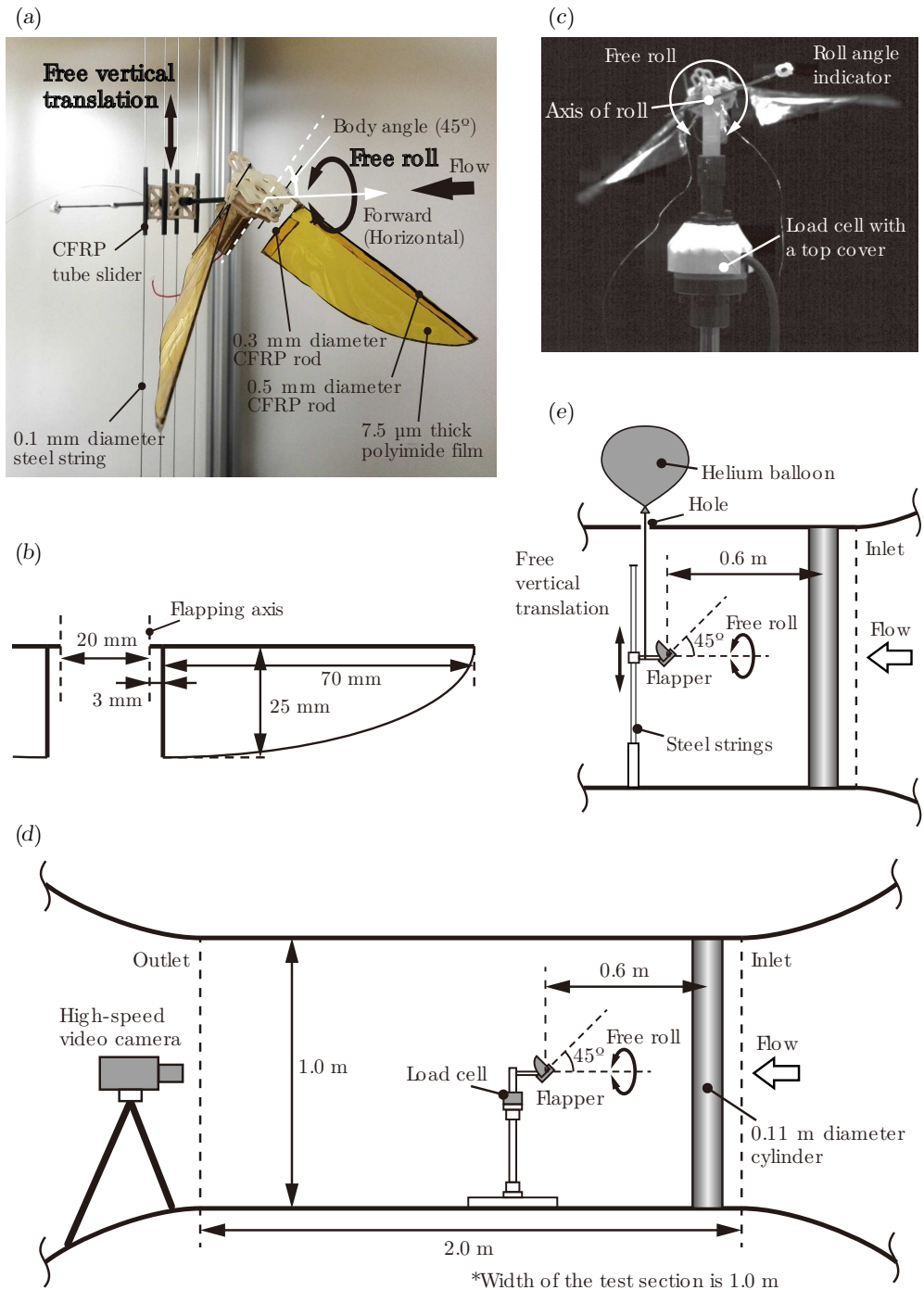}
\caption{\label{fig:flapper_all}
(\textit{a}) Robotic flapper attached on a roll bearing and vertical rail allowing two degrees of freedom (2-DOF).
(\textit{b}) Schematic drawing of the wings, showing the planform and the position of flapping axis.
(\textit{c}) Backside view of the flapper attached to a load cell with unconstrained roll motion (1-DOF).
(\textit{d}) Schematic drawing of the force measurement experiment in a wind tunnel.
The cylinder is installed to produce unsteady inflow, and removed to produce steady inflow conditions on the flapper.
(\textit{e}) Schematic drawing of the 2-DOF flight experiment in a wind tunnel.
The setup permits unconstrained vertical translation and roll rotation of the flapper.}
\end{center}
\end{figure*}

The robotic flapper was attached to an ATI Nano17 Titanium (ATI Industrial Automation, Inc., USA) force sensor through an adapter to measure the produced vertical force (figure~\ref{fig:flapper_all}\textit{c}). Forces were sampled using a NI USB 6343 DAQ board sampling at 1kHz.
To simulate forward flight, the stroke plane angle of the flapper was fixed at $45$~deg with respect to the horizontal. Force measurements on the flapper were taken in two configurations, i.e.,
zero degrees of freedom (0-DOF, fully tethered) and one degree of freedom (1-DOF, tethered but free to roll). In the former, the flapper was rigidly attached to the force sensor through a carbon fiber (CFRP) rod.
For the 1-DOF tests, the flapper was attached to the adapter using
a concentric CFRP tube/rod pair to create a low friction bearing that allows roll.
The axis of rotation was parallel to the horizontal plane and located $6$~mm above the center of mass inducing positive pendulum stability.  Force measurements were taken at three input voltages between $2.5$ and $3.5$~V at $0.5$~V increments. 
The voltage and current drawn by the flapper were sampled at $1$~kHz.
Input power was calculated by multiplying the time-averaged values of the voltage and the current.

Experiments were conducted in a closed return wind tunnel (figure~\ref{fig:flapper_all}\textit{d}) with a $1 \times 1 \times 2$~m test section with a mean wind speed of 3.5~m/s.
This was within the typical range of hummingbird flight speed, from 0 to 12~m/s \cite{Tobalske_etal_2007_jeb}.
In the unimpeded configuration, steady airflow was measured within the test section with turbulence intensity $<2\%$ defined with respect to the mean flow.
Unsteady wind was generated by placing a cylinder with diameter $D=11$~cm near the inlet of the test section at distance $x=0.6$~m in front of the flapper.
This created a von K\'arm\'an vortex street at $Re_D = 35,000$ that induced alternating lateral disturbances on the flapper at the characteristic frequency of $f_{vk}=9$~Hz.

\begin{figure*}
\begin{center}
\includegraphics[scale=0.85]{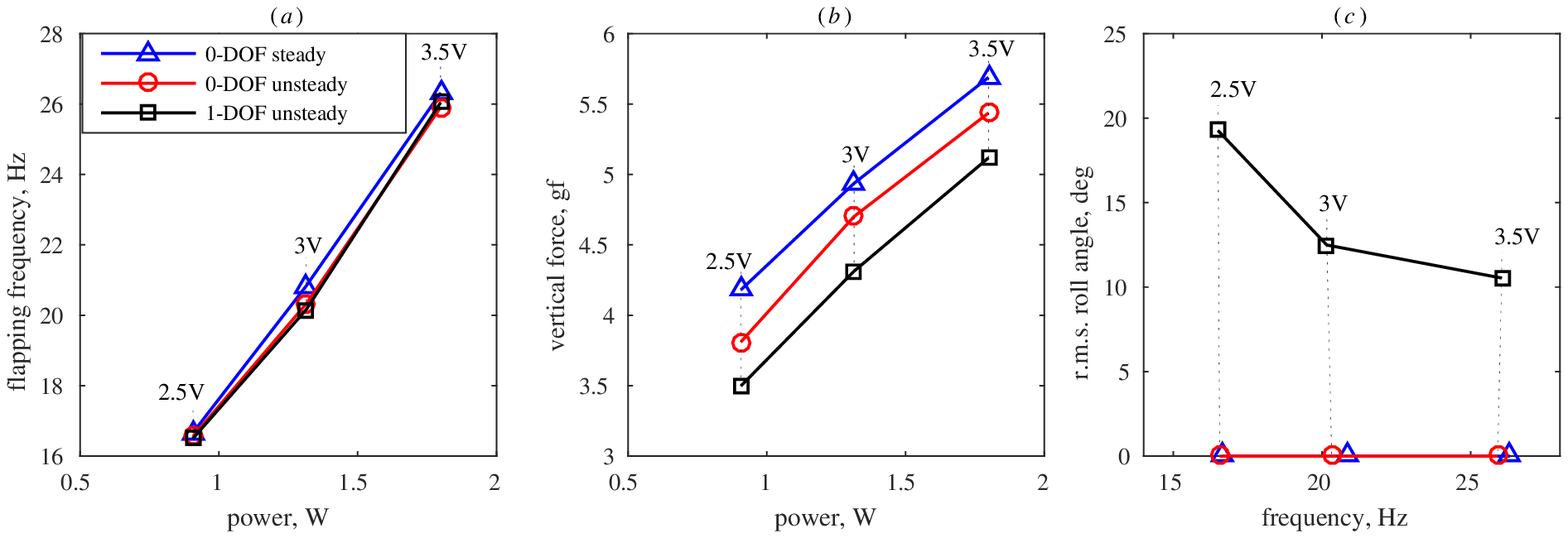}
\caption{\label{fig:force_meas} Results of 0-DOF and 1-DOF robotic flapper experiments: (a) flapping frequency versus power; (b) mean vertical force versus power; (c) r.m.s. roll angle versus flapping frequency. Dotted lines connect points of equal voltage. Vertical force is in gram-force (gf) units.}
\end{center}
\end{figure*}

When the unsteady inflow was generated using a cylinder,
the wind speed of the wind tunnel was increased to compensate for the velocity deficit in the wake.
We used a hand held anemometer (Kanomax Climomaster) to ensure that the time average velocity of 3.5~m/s was maintained at the location of the flapper in all cases.

When the flapper was fully tethered (0-DOF) and when roll freedom was permitted (1-DOF), in both steady and unsteady airflow conditions, increasing voltage led to
monotonic increase of the flapping frequency, and, consequently, to increasing aerodynamic forces and power, see figure~\ref{fig:force_meas}. For a given voltage, the current
was nominally invariant across the three testing conditions, and
the flapping frequency of the robot changed by less that 0.7~Hz (figure~\ref{fig:force_meas}\textit{a}).

The vertical aerodynamic force, figure~\ref{fig:force_meas}(\textit{b}), varied by up to 20\% across the three testing conditions.
For a given voltage (which determined the power), the flapper under fully tethered (0-DOF) condition in steady airflow produced the largest vertical force.
Under fully tethered condition in the unsteady airflow, the flapper produced around 0.29~gf (6\%) less vertical force compared to that in steady winds, on average over the range of voltages tested.
The vertical force produced by the flapper in unsteady airflow when permitted to roll (1-DOF) dropped by additional 0.34~gf (7\%), on average.

The reduction in vertical force between the two 0-DOF tests may be due to two factors.
First, as pointed out in \cite{Fisher_etal_2016_bb} for a fully tethered flapper, its sensitivity to inflow velocity fluctuation increases as the flapping frequency decreases.
The flapping frequency of our robot may be sufficiently low to produce similar effects.
Second, non-uniformity of the mean velocity profile and lateral velocity contribution to dynamic pressure
may affect the time-averaged forces. Analysis of these effects is beyond the scope of our theory, but further experiments investigating the aerodynamic interactions between flapping wings and free stream turbulence will be very useful.

Here we focus on the influence of body roll on vertical force production in unsteady airflow.
Figure~\ref{fig:force_meas}(\textit{c}) shows an decrease in r.m.s. roll,
consistent with the $f^{-1}$ power law derived in section~\ref{sec:roll_dynamics}.
Consequently, using (\ref{eq:theor_fz_deficit}), we estimate the roll-induced vertical force deficit
to be up to 0.2~gf.
Note that, as voltage increases, the absolute force deficit $\Delta \overline{F}_z$
in figure~\ref{fig:force_meas}(\textit{b}) remains approximately constant,
but the relative force deficit $\Delta \overline{F}_z / \overline{F}_{z0}$ decreases,
as expected from the theoretical estimates for sufficiently large $Tu_w$.
As the roll amplitude decreases, reorientation of the net aerodynamic force vector becomes insignificant.
Therefore, with respect to the aerodynamic force magnitude that increases with $f$, the mean vertical force deficit due to body rotations becomes smaller.

\subsection{Semi-restrained flight measurements with free vertical translation and roll (2-DOF)}

To demonstrate the influence of body roll on the weight support, flight measurements using the flapping wing contrivance were conducted.
In the flight measurements, the flapper was attached on a vertical nominally frictionless rail that consisted of four steel strings with diameter 0.1~mm, held in tension. Custom guides were used to attach the flapper to the rail thus permitting uninhibited vertical movement, see figures~\ref{fig:flapper_all}(\textit{a},\textit{e}). 
Similar to the tethered 1-DOF measurements, a CFRP low friction bearing enabled the flapper to roll.

First, the power required for the flapper to hover under steady winds was measured by sequentially increasing the voltage until sustained hovering was noted.
Subsequently, the cylinder was installed at the inlet of the test section and the procedure was repeated
after adjusting the wind tunnel speed to account for the wake velocity deficit.
The power required to hover in unsteady winds was measured.

Instantaneous roll angle of the flapper was obtained from optical tracking using a high-speed video camera (FASTCAM SA-3, Photron Ltd., Japan) placed downstream.
High-speed videography was conducted at 500Hz and a custom code written in MATLAB was used to track the vertical position of the flapper.

\begin{figure*}
\begin{center}
\includegraphics[scale=0.85]{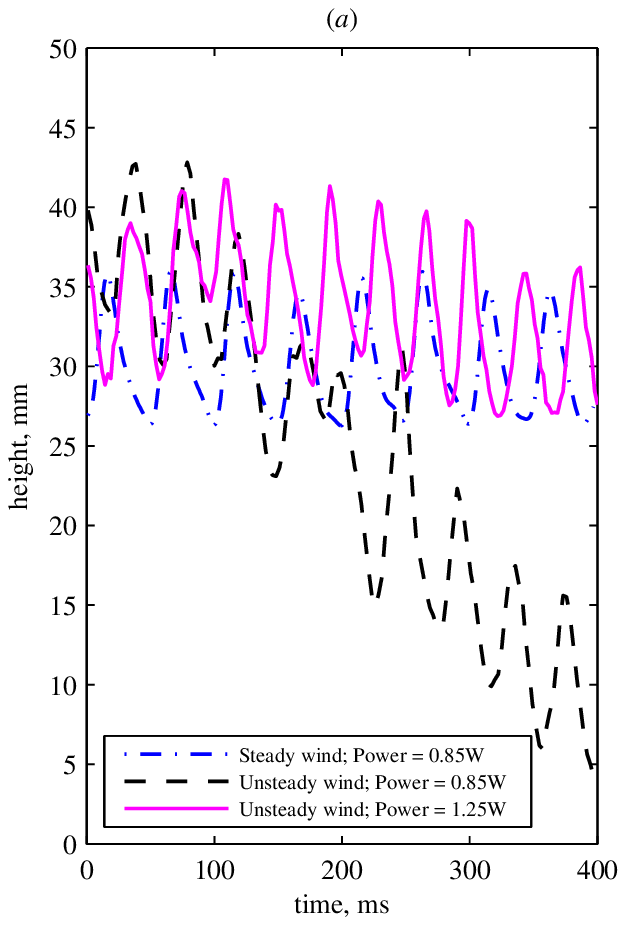}
\includegraphics[scale=0.91]{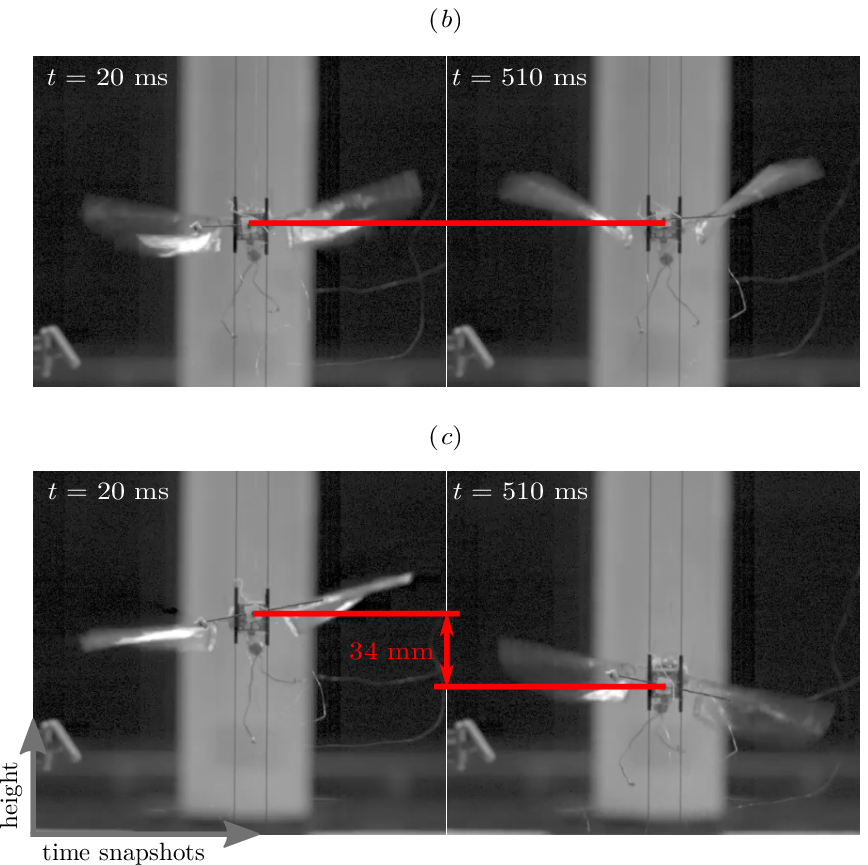}
\caption{\label{fig:2dof_meas} Results of the 2-DOF flight tests. (\textit{a}) Time evolution of the flapper's vertical position in three different flight tests.
(\textit{b}) Two subsequent photographic images of the flapper in level flight in unsteady wind, with the power input of $1.25$~W.
(\textit{c}) Photografic images in descending flight in unsteady wind, with the power input of $0.85$~W.}
\end{center}
\end{figure*}

A decidedly important regime from the point of view of practical MAV applications is level flight.
In steady as well as unsteady inflow conditions, this regime implies that the time average lift should support the weight over a sufficiently long time.
In our 2-DOF flight tests, where the robotic flapper is permitted to move along the vertical axis and rotate along the wind tunnel long axis,
we test the power required to ensure the level flight on steady and unsteady winds.
Compared to steady winds, the flapper rolled considerably in the wake of the cylinder.

The results of the 2-DOF flight tests are summarized in figure~\ref{fig:2dof_meas}.
Panel (a) shows the height of the roll hinge axis above an arbitrarily selected reference.
The height variation of the flapper with the flapping phase was due to the inertia of the wings themselves, and is to be expected for low wing loading flappers.
Butterflies and moths likewise exhibit similar motion due to high wing inertia. For the flapper, level flight in steady wind required power input of 0.85~W,
however the same power was insufficient for maintaining level flight in the unsteady wake past the cylinder, see figures~\ref{fig:2dof_meas}(\textit{b}) and (\textit{c}). In latter case, after the flapper was released,
it descended at rate of 0.1~m/s. When the power input was increases to $1.25$~W in the unsteady wind conditions, sustained level flight was achieved, as shown in figure~\ref{fig:2dof_meas}(\textit{a}).
The increased power requirement for level flight in unsteady winds is consistent with the force measurements which showed that the flapper produces
diminished vertical force when permitted to experience body rotations in unsteady winds (cf. figure~\ref{fig:force_meas}).

\section{Numerical simulations of a bumblebee}\label{sec:numerical_simulation}

To explore the parameter space of flapping flight in unsteady winds that was beyond the testing range of the robotic flapping wing contrivance,
numerical simulations of a bumblebee flight were conducted. The bumblebee model used for this study is essentially the same as in our earlier work on the aerodynamics and flight dynamics of bumblebees in turbulence \cite{Engels_etal_2016_prl,Ravi_etal_2016_srep}. Its brief summary can be found in the supplementary material~S2. 

\begin{figure*}
\begin{center}
\includegraphics[scale=0.9]{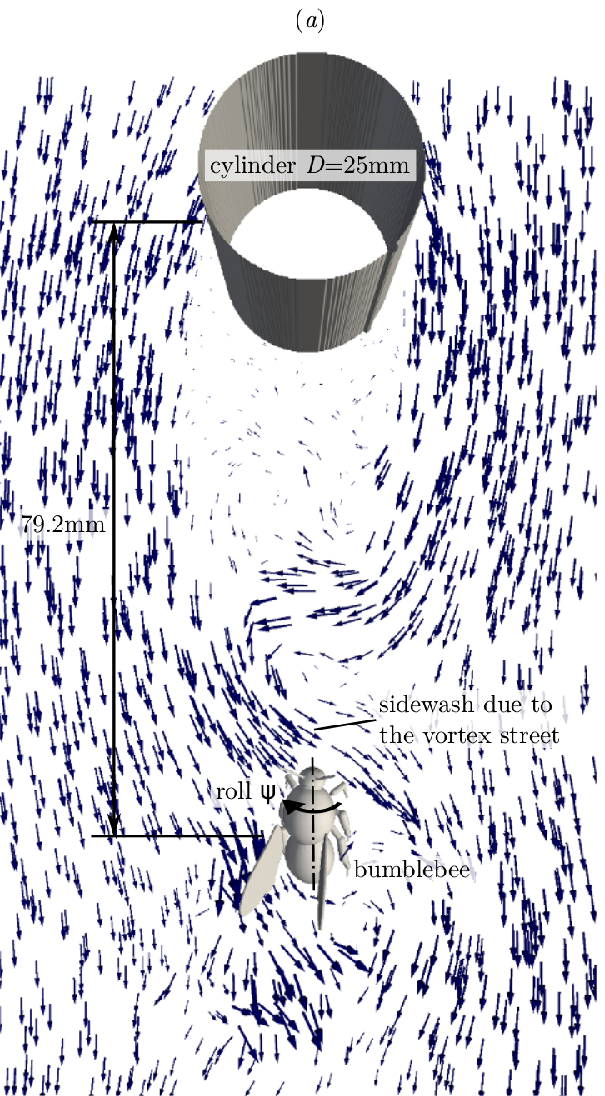}\quad
\includegraphics[scale=0.9]{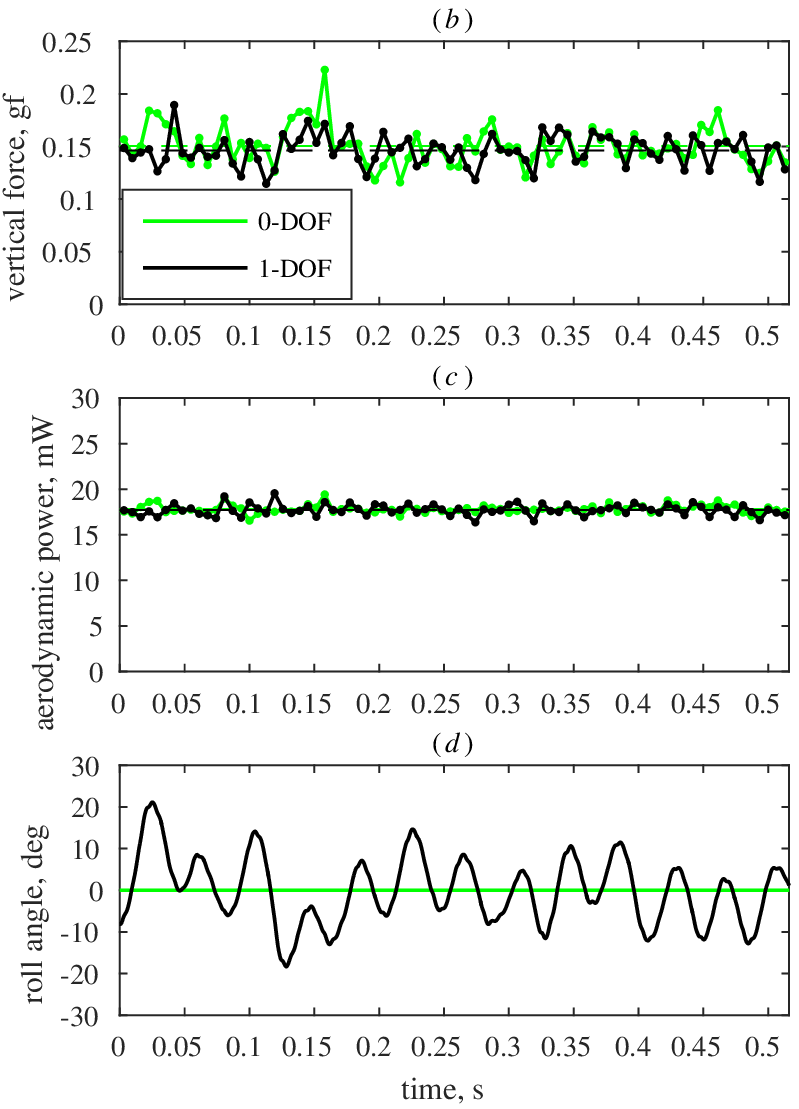}
\caption{\label{fig:num_results} Results of the numerical simulations of bumblebee flight in the wake of a cylinder. (\textit{a}) Visualization of the 1-DOF bumblebee-cylinder system and
velocity field in a horizontal plane passing through the center of mass of the bee. (\textit{b}) Vertical aerodynamic force in the laboratory reference frame, (\textit{c}) Aerodynamic power and (\textit{d}) roll angle, over the 0.516~s time intervals corresponding to the 0-DOF and 1-DOF flight subsequences. Dashed lines show the mean values over the analyzed time interval.}
\end{center}
\end{figure*}

Two simulations of the bumblebee flight in the unsteady wake of the cylinder were conducted under two different imposed conditions, respectively. In the first configuration, the bumblebee model
was fully tethered, i.e., all degrees of freedom were restricted (0-DOF). In the second case, the bumblebee model was free to roll (1-DOF),
i.e., roll rotation about the longitudinal axis of the body varied in time under the action of the external aerodynamic torque.
In both cases, the vertical force with respect to the wind tunnel reference frame, the aerodynamic power, as well as the roll angle in the 1-DOF case, were obtained from the numerical simulations.

The numerical simulation of the bumblebee in fully tethered state continued for 0.645 seconds (100 wingbeats), after which the roll degree of freedom was released and the simulation was restarted
for additional 0.645~s. The von K\'arm\'an street developed after about 0.1~s from the impulsively started flow.
Therefore, only the last 0.516~s (80 wingbeats, 13 vortex shedding periods) of each of the two simulations were analyzed after discarding the transient phase.

The vertical force averaged over every wingbeat is shown in figure~\ref{fig:num_results}(\textit{b}), as a function of time at the middle of each wingbeat.
In both cases, its variation in time induced by the change in wind speed and direction due to the von K\'arm\'an street (see figure~\ref{fig:num_results}\textit{a}) was of order 20\% of the mean value over the analyzed simulation period.
Under the 0-DOF conditions, the mean vertical force was about 150~mgf, while in the 1-DOF case it decreased to 146~mgf, i.e., reduced by 2.7\%.
This is a detectable, but weaker variation than found in the robotic flapper experiments in section~\ref{sec:robotic_flapper_experiments}.
The aerodynamic power showed much less variation, it only reduced by 0.6\%, see figure~\ref{fig:num_results}(\textit{c}).

The roll angle $\psi$, shown in \ref{fig:num_results}(\textit{d}), expectedly remained zero through the 0-DOF flight sequence.
In the 1-DOF case, the bee’s roll angle oscillated almost periodically at the von K\'arm\'an frequency.
The primary disturbance from the von K\'arm\'an street occurred at around $f_{vk}=25$~Hz.
The mean value of $\psi$ calculated over the 0.516~s time interval was very close to zero (-0.02~deg),
and the root mean square was equal to $\psi_{rms} = 7.96$~deg. Since no active control was implemented in the simulations, the observed flight dynamics is only a consequence of passive interactions between the bee model and the von K\'arm\'an street. Ravi \textit{et al.} \cite{Ravi_etal_2016_srep} showed that living bees likewise interact with the von K\'arm\'an street passively and have similar roll oscillations to the numerical bees.
Thus, it is likely that the bees need to increase force production to maintain weight support, which implies additional energetic cost.

\section{Discussion}\label{sec:discussion}

The low inertia of insects, small birds and flapping wing MAVs can
render them susceptible to the adverse effects of wind unsteadiness, including
large variation in body position and orientation during flight. Observations made on insects and birds flying in unsteady winds reveal that animals employ a wide range of flight control strategies to correct for aerial disturbances.
Bees, moths and hummingbirds have been noted to employ both transient and mean variations in wing and body kinematics when
contending with adverse winds \cite{Combes_Dudley_2009_pnas,Ravi_etal_2013_jeb,Ortega-Jimenez_etal_2013_jeb,Ravi_etal_2015_jeb,Ortega-Jimenez_etal_2016_rspb}.
Hummingbirds flying in fully developed turbulent winds increased wingbeat frequency and amplitude,
both of which generally equate of increased force production \cite{Ravi_etal_2015_jeb}.
Hawkmoths increase the wingbeat frequency but may slightly reduce the amplitude \cite{Ortega-Jimenez_etal_2013_jeb}.
While the energetic cost of performing corrective flight manoeuvres is still unclear, the metabolic rate of hummingbirds increased when flying in unsteady winds,
but only when the perturbations and body rotations were high \cite{Ortega-Jimenez_etal_2014_procb}.
This is consistent with our results from the robotic flapper in section~\ref{sec:robotic_flapper_experiments}.

Oscillations in body roll and its energetic overhead are likely to be generic effects to all flapping fliers in unsteady wind, both living organisms and MAVs.  Here we assess the suitability of the functional relationships derived in section~\ref{sec:theoretical_estimates} by calculating quantitative estimates for various biological and robotic systems flying in unsteady winds and comparing the result with
the values obtained in experiments.

The simplest formula of roll amplitude (\ref{eq:theor_roll_amp_approx}) contains two unknown parameters $c_{\tau 1} / \kappa_{fct}$ and $c_{\tau 2} / \kappa_{fct}$.
These parameters depend on the flier's morphology, but from dynamical similarity considerations
it is reasonable to neglect their variability for the purpose of deriving approximate general estimates. Using the measurements from flight experiments on living and artificial systems in von K\'arm\'an streets conducted in this and prior studies, nominal estimates for $c_{\tau 1} / \kappa_{fct}$ and $c_{\tau 2} / \kappa_{fct}$ can be obtained by minimization of the least mean square error with respect to the theoretical estimate of the r.m.s. roll angle (see supplementary material~S3).
This results in the following values with an error of around 4.5~deg:
\begin{equation}
c_{\tau 1} / \kappa_{fct} = 0.267, \quad c_{\tau 2} / \kappa_{fct} = 1.603,
\label{eq:ctau_values}
\end{equation}

\begin{figure*}
\begin{center}
\includegraphics[scale=0.9]{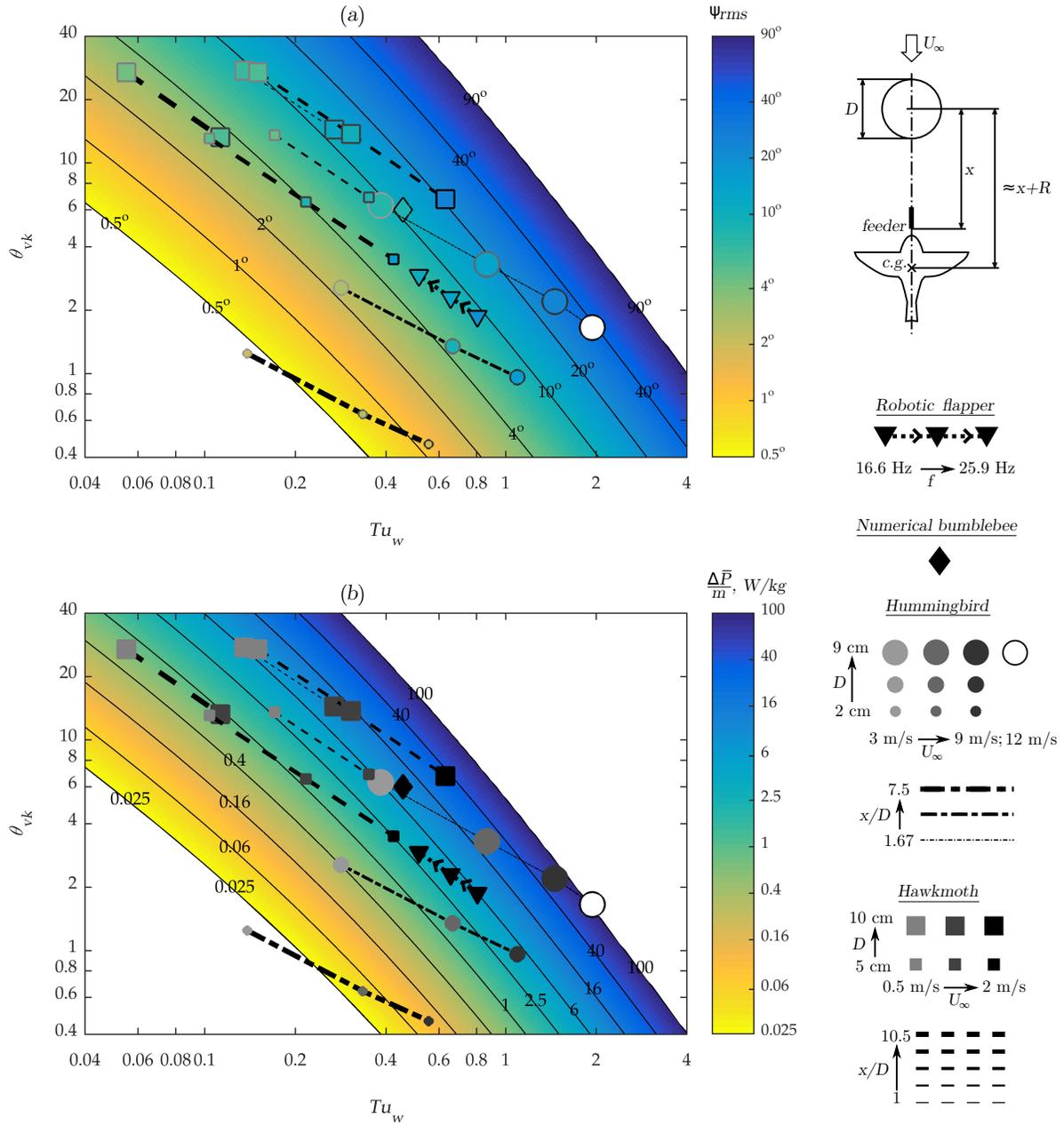}
\caption{\label{fig:diagram_theory} Diagrams of (\textit{a}) r.m.s. roll angle and (\textit{b}) body mass specific mechanical power due to body rotation,
as functions of dimensionless parameters $Tu_w$ and $\theta_{vk}$, as defined by (\ref{eq:eq_turb_int}) and (\ref{eq:thetavk}), respectively.
Isolines and the color fill between them correspond to theoretical estimates.
Markers of different shapes correspond to different sets of experiments. The color of the markers in figure~(\textit{a}) corresponds
to the values of $\psi_{rms}$ measured in the experiments. The size of circles and squares encodes the cylinder diameter $D$ in the hummingbird and hawkmoth experiments
\cite{Ortega-Jimenez_etal_2014_procb} and \cite{Ortega-Jimenez_etal_2013_jeb}, respectively. The level of gray encodes inflow velocity $U_\infty$.
Multiple data points that correspond to flights at the same relative distance from the cylinder $x/D$ are connected with dashed or dash-dot lines, for the ease of visualization.
The line thickness encodes the value of $x/D$ in each set of experiments.
The diagrams shows that, if a point is situated near the top-right corner of the domain,
the corresponding flight condition produces larger r.m.s. roll and larger body mass specific mechanical power than
for a point situated near the bottom-left corner.}
\end{center}
\end{figure*}

With the values of aerodynamic coefficients defined in (\ref{eq:ctau_values}),
equations~(\ref{eq:theor_roll_amp_approx}) and (\ref{eq:theor_roll_rms_approx}) provide an estimate of the r.m.s. roll angle experienced by arbitrary flappers
due to the flow unsteadiness. It is visualised in figure~\ref{fig:diagram_theory}(\textit{a}) for a range of $Tu_w$ and $\theta_{vk}$. Due to the inherent non-linear relationship between the $Tu_w$ and $\psi_{rms}$  that is mediated by flapper properties, the slope of the isolines changes between the two asymptotic regimes, i.e., $Tu_w$ small and $Tu_w$ large asymptotes. In both regimes, however, $\psi_{rms}$ of a flapper increases with $Tu_w$, because larger r.m.s. lateral velocity leads to larger roll over a fixed period of time.
Concomitantly, $\psi_{rms}$ is also increased with $\theta_{vk}$. This occurs because longer periods of oscillation allow for larger roll amplitude, if the forcing amplitude is held constant
($\Psi \propto \tau_a' / f_{vk}^2$).

Insights into the relationships between the different parameters that influence $\psi_{rms}$ for a flapper in different airflow conditions can be obtained by plotting measurements of $\psi_{rms}$ for different flapping wing systems alongside the theoretical predictions derived using the above values for  $c_{\tau 1} / \kappa_{fct}$ and $c_{\tau 2} / \kappa_{fct}$ (figure~\ref{fig:diagram_theory}\textit{a}). In the case of the experiments and numerical simulations conducted here, the theoretical estimates under predict the $\psi_{rms}$ of the robotic flapper while over predicting the $\psi_{rms}$ for the numerical bumblebee by around 6~deg and 3~deg on average, respectively. This could be a result of the
model coefficients being optimised to all fliers rather than specifically adjusted to the robotic flapper or the numerical bumblebee. Importantly, the trend of decreasing variations in roll angle with increasing flapping frequency is consistent between the experiment and the theory.

Similarly, flight measurements made on hawkmoths  \cite{Ortega-Jimenez_etal_2014_procb} and hummingbirds  \cite{Ortega-Jimenez_etal_2013_jeb} can also be compared with theoretical estimates derived here. Both hawkmoth and hummingbirds have their wing length and flapping frequency of the same order of magnitude as the parameters of our flapper, which justifies the use of the same value of the model coefficients.  In the hummingbird experiments \cite{Ortega-Jimenez_etal_2014_procb}, among the cases tested, the largest value of $\psi_{rms}$ corresponds to the largest cylinder diameter $D$ and the largest velocity $U_\infty$,
with the theoretical estimate being equal to 29~deg, to be compared the measured value of 23~deg.
The roll amplitude becomes smaller as $D$ and $U_\infty$ decrease.
As the cylinder diameter $D$ decreases, it results in an increase in vortex shedding frequency $f_{vk}$, i.e., smaller $\theta_{vk}$.
In addition, with the distance $x$ being invariant (in the case of  \cite{Ortega-Jimenez_etal_2014_procb}), smaller $D$ implies greater $x/D$,
i.e., smaller $W'$ and $Tu_w$. These two effects sum up to reduce $\psi_{rms}$.
As per $U_\infty$, its decrease makes $W'$ and $Tu_w$ smaller, but, on the other hand,
slower speed means lower vortex shedding frequency $f_{vk}$, i.e., greater $\theta_{vk}$.
In that case, the effects of $Tu_w$ and $\theta_{vk}$ are opposite. The roll amplitude in (\ref{eq:theor_roll_amp_approx}),
depends on $Tu_w^2$, which dominates in the hummingbird regime and makes $\psi_{rms}$ decrease as $U_\infty$ decreases.
For the smallest cylinder, the effect of flow unsteadiness is negligible in that case, in agreement with the theory.
The hawkmoth experiment data points \cite{Ortega-Jimenez_etal_2014_procb} populate a different subset of the parameter space in figure~\ref{fig:diagram_theory}(\textit{a}). They reveal that
even the largest cylinder can have little effect if the insect is far enough from it,
which can be explained by the decay of $W'$, i.e., small $Tu_w$.

The energetic cost due to body rotations in unsteady flows can be estimated by applying the theoretical estimate (\ref{eq:theor_power})
to  $\psi_{rms}$ estimated for the test cases presented in figure~\ref{fig:diagram_theory}(\textit{a}).
We assume constant gradient
$\partial \overline{P} / \partial \overline{F}_{z} = 0.5$~W/gf $= 51$~W/N. This estimate is obtained from
the robotic flapper measurements shown in figure~\ref{fig:force_meas}(\textit{b}),
and it seems adequate for both hawkmoth and hummingbird, which are of similar size.
Moreover, if mass specific power is of the same order of magnitude for all species regardless of their size,
so is the gradient $\partial \overline{P} / \partial \overline{F}_{z}$.

Based on this approximation, figure~\ref{fig:diagram_theory}(\textit{b})
shows the additional body mass specific mechanical power required for flight in vortex streets.
The isolines of $\Delta \overline{P}/m$ are colored according to the theoretical estimate.
Markers show how different experiments are positioned on the diagram,
such as to allow approximate evaluation of each case. Similar to  $\psi_{rms}$,
the mass specific mechanical power for both the robotic flapper and living systems also increases with increase wind unsteadiness (figure~\ref{fig:diagram_theory}\textit{b}).

In the most challenging condition tested in \cite{Ortega-Jimenez_etal_2014_procb} (case $D=9$~cm, $U_\infty=9$~m/s), the power increment for hummingbirds $\Delta \overline{P}/m$ amounts to 67~W/kg.
That would add up to about 34~W/kg required for forward flight under steady conditions \cite{Song_etal_2016_rsos}.
Even though this value is large, it is below the maximum mass-specific aerodynamic power produced during load-lifting \cite{Altshuler_etal_2010_jeb}.
In the next two severe cases ($D=9$~cm, $U_\infty=6$~m/s and $3$~m/s), we obtain $\Delta \overline{P}/m=24$~W/kg and $6$~W/kg, respectively.
The power increment is less than 5~W/kg in all of the remaining hummingbird cases.
This trend is in agreement with the metabolic rate measurements in \cite{Ortega-Jimenez_etal_2014_procb}.
Metabolic rates were similar for all airspeeds at both the control (no cylinder) and the medium cylinder experiments, but significantly larger in the large cylinder experiments.
In the hawkmoth experiments, the theory suggests that $\Delta \overline{P}/m$ varies between 0.6 and 34~W/kg.

While it is desirable for a flapper to minimize body rotations experienced while flying in unsteady winds and concomitantly reduce energetic cost,
the relationships derived in section~\ref{sec:roll_dynamics} and figure~\ref{fig:diagram_theory}(\textit{a}) suggest that $\psi_{rms}$ is dependant on $Tu_w$, $\theta_{vk}$.
From figure~\ref{fig:diagram_theory}(\textit{a}) it would be ideal for a flapper to reduce $Tu_w$ and $\theta_{vk}$ but as they are not mutually independent,
varying the flapping frequency $f$ as a compensatory response can lead to movement along an isoline resulting in limited overall reduction in $\psi_{rms}$ or energetic cost.

Optimizing the torque coefficients $c_{\tau 1}$ and $c_{\tau 2}$ may offer some potential for minimizing the roll-induced energetic costs.
Real systems are very dynamic and their interaction with unsteady winds can be nonlinear - dependant on a number of factors including passive stability, flexibility, control over various kinematic parameters, etc.
All these effects would directly influence  $c_{\tau 1} / \kappa_{fct}$ and $c_{\tau 2} / \kappa_{fct}$ that would in turn influence the relationship between $Tu_w$, $\theta_{vk}$ and $\psi_{rms}$.  In this analysis we assume the abstracted case of the von K\'arm\'an street as representative of unsteady flow,
airflow in the outdoor environment is generally fully turbulent that will impose broadband perturbations.
In such conditions, the dynamic interaction between airflow and flapping wings needs further considerations,
since our analysis assumes quasi-steady time periodic interactions.

For the precise estimation of magnitude of body rotations experienced and added energetic costs of flight in unsteady winds,
the specific static and dynamic properties of the flapper must be accounted for in calculating the various coefficients.
Therefore isolines presented in figure~\ref{fig:diagram_theory} and the subsequent deductions need to be treated with some
level of caution since they assume generic values for the coefficients.
However, the generic values assumed here provide quantitative evidence to the added energetic costs of flapping flight in unsteady adverse winds.

In our robotic flapper experiments, to ensure mean weight support while being destabilized by the wind, the flapping frequency $f$ was increased.
In the case of biological fliers, apart from changing flapping frequency they have been shown to implement diverse mechanisms to compensate for reduction in vertical force
due to body rotation,
such as hummingbirds flying in turbulent winds tend to increase the mean fan angle that not only aids pitch stabilisation
but also contributes to aerodynamic force production \cite{Ravi_etal_2015_jeb}. However the birds also experienced increase in body drag that will likely increase metabolic rate.
Hawkmoths were found to increase net aerodynamic force by flapping with
larger amplitude, in elevation and sweep, during voluntary lateral manoeuvres \cite{Greeter_Hedrick_2016_jeb}.
For the case of artificial flapping wing systems further studies are necessary to identify optimal mechanisms for
flight control and force production through which added energetic costs incurred due to body rotations are minimal.

\section{Conclusions}\label{sec:conclusions}

We hypothesised that the large body rotations experienced by small-scale
flapping fliers in unsteady airflows can result in a cumulative reduction
in vertical force due to the reorientation of the aerodynamic force vector
during body rotations.
We introduced dimensionless parameters $Tu_w$ and $\theta_{vk}$
that characterize, respectively, the intensity and the frequency of aerodynamic
perturbations relative to the wing flapping motion.
We derived a functional relationship between these parameters and the body roll amplitude (\ref{eq:theor_roll_amp_approx}),
and estimated the added mechanical power requirement (\ref{eq:theor_power}).

We tested the theory under two
conditions, i.e.,
when the flapping time scale is of the same order as
the disturbance
and
when the time scale of flapping is much higher than the
aerial disturbance.
First, experiments with a miniature robotic flapper, where the disturbance was of
the order of flapping period, revealed that the mean vertical force was
lower when the flapper was permitted to roll as compared to fully tethered
conditions. In semi-free flight conditions, compared to steady wind
conditions, we found that the flapper rolled considerably in unsteady winds and required
higher input power to maintain altitude. The rotations experienced by the
robotic flapper also decreased monotonically with increasing wingbeat
frequency $f$, in agreement with the theory.
Second, high fidelity numerical simulation revealed that, at the
scale of bumblebees where the flapping frequency is higher than the disturbance frequency, unsteady winds can induce large body
rotations that can also translate to a deficit in the vertical force.

Finally, we compared the theoretical findings with the published
data from experiments with animals flying in unsteady wakes
behind vertical cylinders. The proposed theory explained all trends
of the roll angle with respect to the diameter of the cylinder $D$,
the distance from the animal to the cylinder $x$,
and the inflow velocity $U_\infty$.
Theoretical estimates of the added mechanical power
were found to be consistent with the available data on metabolic rates
in hummingbirds.

Strategies employed by insects and hummingbirds to compensate for the vertical force deficit due to body rotations
include increase of flapping frequency and/or stroke amplitude.
Such compensatory controls require additional mechanical power,
which can be substantial if the roll amplitude is large.
Thus, aerial locomotion in complex airflows comes with the increased energetic demands.

~

This work was granted access to	the
HPC resources of IDRIS (Institut du Développement
et des Ressources en Informatique Scientifique) under	
project number i20152a1664. SR and DK
gratefully acknowledge the financial support from the JSPS (Japan Society for the Promotion of Science)
Postdoctoral Fellowship, JSPS KAKENHI No. 15F15718 and 15F15061, respectively.
HL was partly supported by the JSPS KAKENHI No. 24120007 for Scientific
Research on Innovative Areas.
SR would also like to thank the Alexander von Humboldt Foundation for financial support.
TE, KS, and JS thank the French-­‐German University for financial support,
and acknowledge funding in the French-German ANR/DFG project AIFIT (grant ANR 15-CE40-0019).

\bibliographystyle{vancouver}
\bibliography{bibliography}

\end{document}